\documentclass[pdflatex,sn-nature]{sn-jnl}% Style for submissions to Nature Portfolio journals
\usepackage{natbib}
\usepackage{graphicx}%
\usepackage{multirow}%
\usepackage{amsmath,amssymb,amsfonts}%
\usepackage{amsthm}%
\usepackage{mathrsfs}%
\usepackage[title]{appendix}%
\usepackage{xcolor}%
\usepackage{textcomp}%
\usepackage{manyfoot}%
\usepackage{booktabs}%
\usepackage{algorithm}%
\usepackage{algorithmicx}%
\usepackage{algpseudocode}%
\usepackage{listings}%

\newcommand{\araa}{Annu. Rev. Astron. Astrophys.}   % Annual Review of Astron and Astrophys
\newcommand{\aj}{Astron. J.}   % Astronomical Journal
\newcommand{\apj}{Astrophys. J.}   % Astrophysical Journal
\newcommand{\apjl}{Astrophys. J. Lett.}   % Astrophysical Journal, Letters
\newcommand{\aap}{Astron. Astrophys.}   % Astronomy and Astrophysics
\newcommand{\jcap}{J. Cosmol. Astropart. Phys.}   % Journal of Cosmology and Astroparticle Physics
\newcommand{\mnras}{Mon. Not. R. Astron. Soc.}   % Monthly Notices of the RAS
\newcommand{\prl}{Phys. Rev. Lett.}   % Physical Review Letters
\newcommand{\pasj}{Publ. Astron. Soc. Jpn}   % Publications of the Astron. Soc. of Japan (note no full stop following Jpn)
\theoremstyle{thmstyleone}%
\theoremstyle{thmstyletwo}%

\theoremstyle{thmstylethree}%

\begin{document}

\title[Article Title]{Looking beyond lambda}


\author*[1]{\fnm{Alexie} \sur{Leauthaud}}\email{alexie@ucsc.edu}

\author*[2,3]{\fnm{Adam} \sur{Riess}}\email{ariess@stsci.edu}



\affil[1]{\orgdiv{Department of Astronomy and Astrophysics}, \orgname{University of
California, Santa Cruz}, \orgaddress{\street{1156 High Street}, \city{Santa Cruz}, \postcode{95064}, \state{CA}, \country{USA}}}

\affil[2]{\orgname{Space Telescope Science Institute}, \orgaddress{\street{3700 San Martin Drive}, \city{Baltimore}, \postcode{21218}, \state{MD}, \country{USA}}}

\affil[3]{\orgdiv{Department of Physics and Astronomy}, \orgname{Johns Hopkins University}, \orgaddress{\city{Baltimore}, \postcode{21218}, \state{MD}, \country{USA}}}


\abstract{Widening cracks are appearing in the $\Lambda$CDM model and it is becoming increasingly clear that the standard cosmological model struggles to describe the full expansion history of the Universe as revealed by the Cosmic Microwave Background, Baryon Acoustic Oscillation measurements, and locally calibrated Type Ia supernovae. Taken at face value, recent results suggest a dark sector that may be more complex than commonly assumed.  We must prepare for the possibility of moving beyond the $\Lambda$CDM era, where merely testing $w=-1$ is no longer sufficient, and embrace the challenge of unraveling the physics of dark matter, dark energy and gravity on cosmic scales. Guided by increasingly robust data—secured through considerable investment—we should pursue deeper understanding while being open to complexity in the dark sector, rather than settling for the simplest phenomenology. New data from new facilities and a  new dark energy task force could help illuminate the path forward while changes to our scientific practices will be essential to navigate the potentially rocky road ahead.}

%\keywords{keyword1, Keyword2, Keyword3, Keyword4}



\maketitle

\section{Introduction}\label{intro}

One of the greatest success stories in Cosmology has been the development of the $\Lambda$CDM model that describes the key ingredients of our Universe and how they evolve with time. Since the discovery of the late-time acceleration of the universe in 1998 \cite{Riess1998, Perlmutter_1999}, the $\Lambda$CDM model has become the cornerstone of modern cosmology and has withstood rigorous testing from increasingly large data-sets and sophisticated analysis techniques. Remarkably, the model stood up to 4-5 orders of magnitude reduction in its allowed parameter space produced by space-based Cosmic Microwave Background (CMB) experiments.  But while $\Lambda$CDM is both elegant and simple compared to many alternatives, it is also maddeningly elusive. With only 6 free parameters, and a few other core components (the laws of general relativity, a spatially flat and homogeneous universe, known particles, and gaussian initial conditions), $\Lambda$CDM is able to fit an astonishing amount of data, including the exquisitely measured CMB power spectrum (with a fractional error of only $\sim$10$^{-4}$, the angular size of the acoustic peaks in the CMB is one of the most well measured quantities in astrophysics!), long stretches of cosmic expansion history and much of general astrophysics.   However, despite this tremendous success, $\Lambda$CDM fails to explain in any detail the physics of both dark energy and dark matter which comprise $\sim$ 95\% of the Universe today. In this perspective piece, we discuss cracks appearing in the $\Lambda$CDM model and how the cosmology community might prepare for a potential post-$\Lambda$ era.

\section{The dark sector}\label{sec1}

The $\Lambda$CDM model fits a Universe with 70\% of its energy density today in the form of vacuum-energy, represented by the cosmological constant $\Lambda$, 25\% in the form of Cold Dark Matter (CDM), with the remaining 5\% being in the form of regular matter (``baryons"). Both dark matter and dark energy point to physics beyond the Standard Model of Particle Physics -- understanding these two components has the potential to revolutionize physics \cite{Murayama_2023}.

The search for dark matter is a needle in a haystack problem: we have a number of compelling candidates, but beyond determining that dark matter is ``Cold" (it clumps gravitionally) and ruling out some regions of parameter space (e.g., cross section vs. dark matter mass), we are still pretty much in the dark. Testing for specific models can be very difficult, and each candidate comes with its own set of often complicated signatures to be compared against data. Most analyses assume that dark matter is one single particle, but this is merely out of convenience and simplicity, not because we have theoretical reasons to believe that dark matter is simple. In reality, there could be several distinct types of dark matter, or even worse, ``dark matter" may be a manifestation of a whole hidden sector of dark particles and forces \cite{Arbey_2021}.

When it comes to dark energy, $\Lambda$CDM assumes there is a constant equation of state, with the internal pressure and energy density having a fixed ratio of $w=P/(\rho c^2)=-1$. In Einstein's General Relativity, the internal pressure of vacuum energy carries additional gravity, in this case, negative (i.e. repulsive). If energy is an eternal, static property of the vacuum of space, then cosmic expansion eventually dilutes the attractive gravity of matter (dark or otherwise) until the low-level repulsive gravity of the vacuum wins and accelerates the expansion. In contrast, ``dynamical dark energy" (a phenomenological model) could mimic vacuum energy, being the potential energy of a slowly changing (minimally kinetic) scalar field and thus could have an equation of state $w$ that varies with time. Unlike dark matter, we have few compelling theoretical models for dark energy. But similarly, nothing guarantees that dark energy is even as simple as a single-field vacuum type of energy. Given similarities between dark energy and inflation, it is not unreasonable to assume that there are multiple fields driving cosmic expansion, perhaps sprinkled throughout the history of cosmic expansion. Other more complex scenarios include the possibility of dark energy interacting with dark matter, or that General Relativity breaks down on cosmological scales which have only recently entered our horizon. 

Recent astronomical observations suggest we may be witnessing the demise of $\Lambda$CDM. While this is an exciting prospect, we also need to face the sobering possibility that $\Lambda$CDM might have anchored us to an assumption of simplicity; the real, dark universe could be far more complex. While it is said that ``extraordinary claims require extraordinary evidence'', it is no longer clear which possibility may be considered the extraordinary one--the failure of a powerful, standard model or the perfection of a phenomenological model that lacks compelling, physical explanation. Lacking a firm physical footing, we must follow the data.  In the following, we suggest that current observations may be pointing us beyond the model and that we may further benefit from profound changes to the scientific structures in which we operate to decipher the dark sector.

\section{Chasing lambda}\label{section2}

In 2006, the Astronomy and Astrophysics Advisory Committee (AAAC) and the High Energy Physics Advisory Panel (HEPAP) formed a subcommittee to advise NASA and NSF on the future of dark energy research. The resulting landmark report of the dark energy task force \cite{albrecht_report_2006} laid out a road map for dark energy research. This roadmap labeled Astrophysical surveys according to their potential to constrain $w(z)$, the dark energy equation of state as a function of time, with ``Stage III" surveys being more constraining than ``Stage II" surveys.

Around this time, an ongoing debate in the Astronomy and High-Energy physics communities centered around the question of whether or not to invest significant resources in Astronomy experiments to probe dark energy. On one side of the argument was the notion that the pursuit of $\Lambda$ might not pay off and that decades of efforts and billions of dollars could end up with little more to show that $w=-1.000...$. The fear was that no hints as to the nature of dark energy would be found and that the opportunity cost to general astrophysics would be large \cite{white2007}. Others took the view that a happy union of Astronomy and High-Energy physics might hit a jack-pot discovery that would reveal the nature of dark energy \cite{kolb2007}.

The bets were placed, the die was cast, and the chase for $\Lambda$ took off. We stand today at the dawn of Stage-IV surveys. While Stage-II and Stage-III surveys largely confirmed $\Lambda$CDM, widening cracks have appeared as Stage-IV surveys have come online. Although some could ultimately prove to be systematics, or statistical fluctuations, there are now enough discrepancies between different data sets that we need to seriously consider that the beyond-$\Lambda$CDM era may be just around the corner.  Upcoming Stage IV (and ultimately Stage V) experiments could very well be operating in the beyond-$\Lambda$CDM era when testing $w(z)=-1$ is no longer frontier science.

\section{Tensions in the Hubble constant}\label{section3}

If $\Lambda$CDM is indeed breaking down, the first and most persistent fracture is the Hubble tension — a decade-long, $\geq 5\sigma$ discrepancy between the locally measured value of the Hubble constant and the value predicted by $\Lambda$CDM when anchored to CMB observations \cite{riess_comprehensive_2021}. Recent results from high angular resolution, ground-based CMB data from the South Pole Telescope (SPT) and the Atacama Cosmology Telescope (ACT) provide an even lower predicted value of $66.59 \pm 0.46$ km s$^{-1}$ Mpc$^{-1}$ which would raise the tension to $>$ 6 $\sigma$ \cite{spt25}.  Early on, this mismatch could be dismissed as the result of an unknown systematic error in one of the measurements. However, years of rigorous scrutiny and independent replication on both sides - from HST and \textit{Planck} to JWST, ESA Gaia, ACT, SPT and DESI - have failed to find any measurement errors while further reinforcing the conflict \cite{verde2024}. Specifically, recent measurements now show a concordance of distance measurements to SN Ia hosts (used to calibrate SN Ia luminosities) between HST and JWST and between Cepheids and TRGB \cite{Riess:2024,Freedman:2025}, with differences attributable to the membership of local SN samples (and resulting measure of $H_0$) reducing as their size increases and their statistics improve \cite{Li:2025}.  Replacing SN Ia with other local measurements gives similar results \cite{Jensen:2025,TD:2025,Scolnic:2024} so that local $H_0$ tension is not dependent on any one tool, team or technique. There is still more observational work to be done but it appears likely that this will refine the size of the difference rather than reveal a confluence of independent systematic errors in multiple measures.
Although potential resolutions may lie within modifications to the dark sector, the precise nature of such new physics remains unclear.

\section{Not only dynamic but also phantom}\label{section4}

Recent results from the Dark Energy Spectroscopic Instrument \citep[DESI,][]{levi2013desiexperimentwhitepapersnowmass,desicollaboration2016desiexperimentisciencetargeting} have dealt $\Lambda$CDM another blow and complicate the picture even further \cite[][]{desi_collaboration_desi_2025-1,desi_collaboration_desi_2025}. With 5000 multi-fiber robotic positioners, DESI is the premier multi-object spectrograph and the first of the Stage IV surveys to come online. The recent DESI Baryon Acoustic Oscillations (BAO) analysis uses three years worth of data and covers the redshift range $0.1<z<3.5$ (Figure 1). When combined with \textit{Planck} and Type Ia supernovae (SNe), the DESI$+$ data disfavor $\Lambda$CDM at the 2.8 to 4.2$\sigma$ level depending on which supernovae sample is used. It is important to underscore the fact that the DESI$+$ results have grown in strength after a similar finding in 2024. They do not depend on any single DESI BAO data point and BAO measurements are widely believed to be robust to most kinds of systematics.  

At the heart of the DESI$+$ result is a growing inconsistency between the expansion history predicted by $\Lambda$CDM compared to measurements from Type Ia supernovae and BAO below $z\sim3$ and the CMB at $z\sim 1100$. As shown in \cite{desi_collaboration_desi_2025}, when the full DESI DR2 data are employed, there is a mis-match between the expansion history predicted by $\Lambda$CDM and the expansion history measured from BAO+CMB+SNe (with the significance of the effect depending on which specific combination of SNe and CMB data are used, e.g., \cite[][]{desiact} and \cite[][]{spt25}). Instead, a ``dynamical dark energy" with a strongly varying equation of state is favored to jointly describe all three data-sets (Figure 2). Here, the term ``dynamical dark energy" is used to describe a generic dark energy for which $w$ varies with time, not a specific physical model. A strongly varying equation of state in itself is astonishing, but the results seem to further suggest that dark energy went through a phantom phase (meaning that $w<-1$, see Figure 2) which is more surprising still because physical solutions usually require $w\ge-1$ \citep{Caldwell_2002}. However, we note that the evidence for the crossing of the phantom divide is not as strong as the evidence for dynamical dark energy at late times \cite[][]{desi_collaboration_desi_2025,lodha_extended_2025}.

\begin{figure}[h]
\centering
\includegraphics[width=0.9\textwidth]{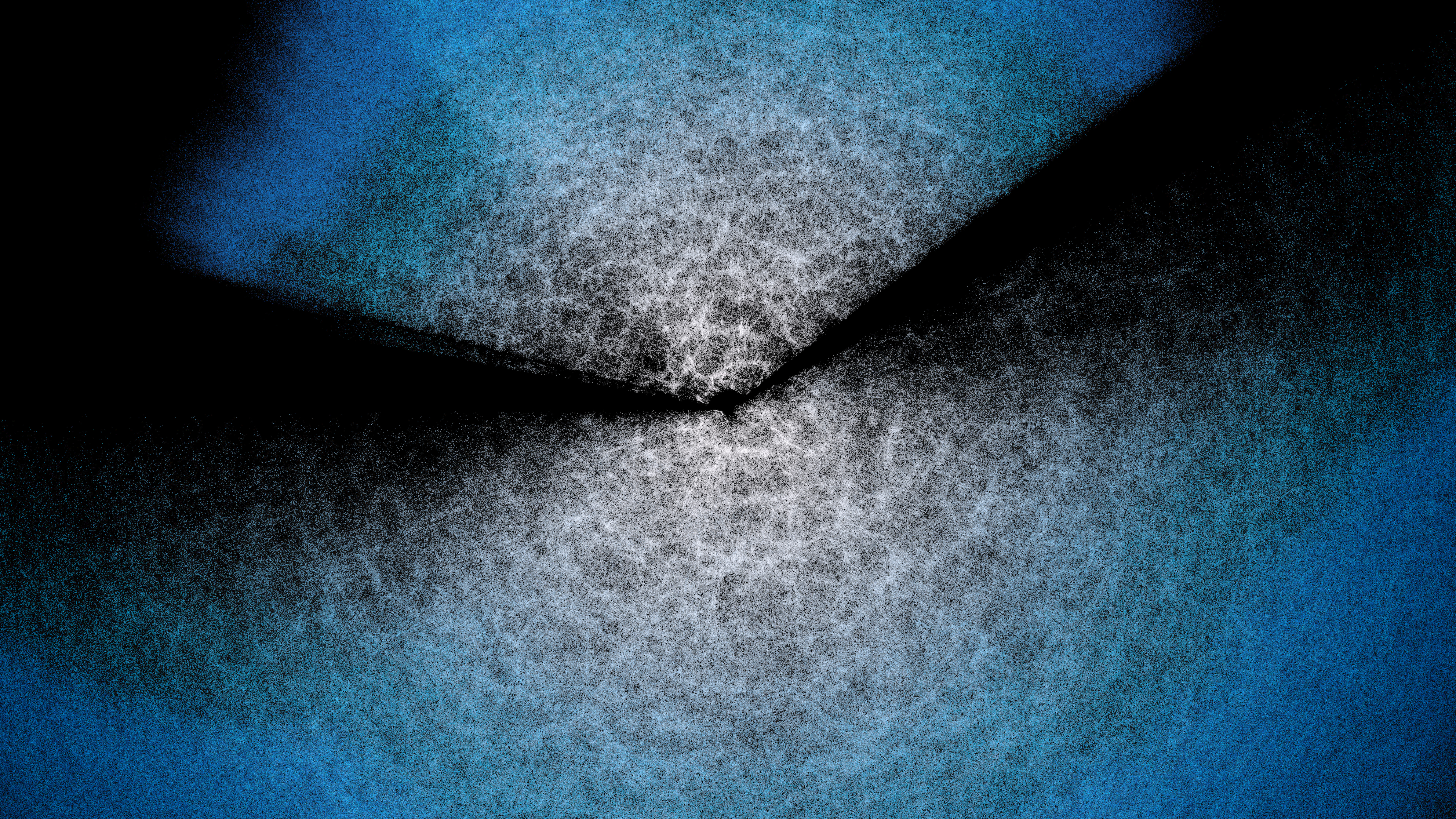}
\caption{\textbf{3D map of the Universe created by DESI}. Earth is at the center in this image, and every dot is a galaxy. Color represents redshift from $z=0$ (center) to $z=4$ (blue outskirts). DESI uses the 3D positions of galaxies to measure the expansion history via Baryon Acoustic Oscillations. 
Credit: DESI collaboration and KPNO/NOIRLab/NSF/AURA/R. Proctor.
}\label{fig1}
\end{figure}

\begin{figure}[h]
\centering
\includegraphics[width=0.9\textwidth]{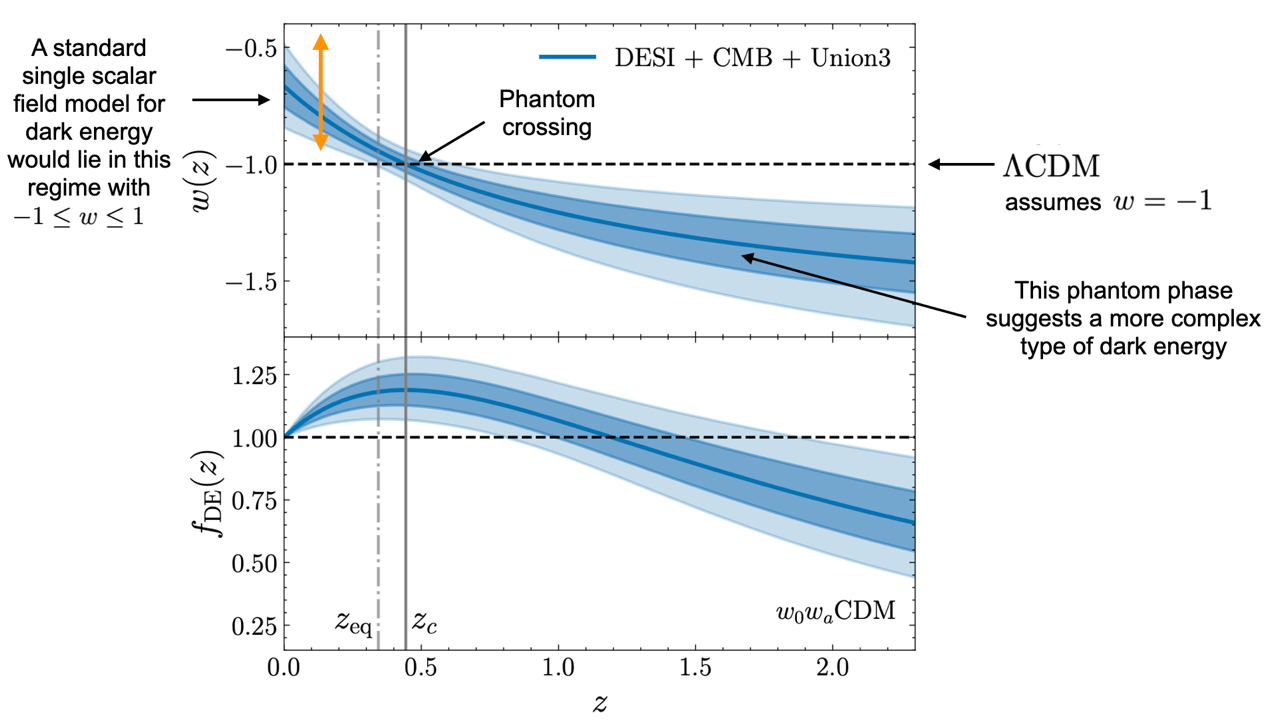}
\caption{\textbf{Dark energy results from DESI.} Annotated version of figure from \cite{lodha_extended_2025} showing equation of state $w(z)$ and dark energy density ($f_{\rm DE}(z)=\rho_{\rm DE}(z)/\rho_{\rm DE,0}$) using the w$_0$w$_a$ parametrization where $w(z)=w_0+w_a\frac{z}{1+z}$. The solid and dashed-dotted vertical lines indicate the phantom-crossing and dark energy-matter equality redshifts, respectively. Shading represents 68\% and 95\% confidence intervals.}\label{fig3}
\end{figure}

\section{A universe that never ceases to surprise}\label{section5}

The DESI$+$ results are surprising for several different reasons \cite{lodha_extended_2025, linder_interpreting_2024}. Taking the DESI$+$ results at face value, they suggest that: a) $w$ is not a constant, b) dark energy has a phantom phase of $w<-1$ at $z>0.5$ and crossed the phantom divide around $z \sim 0.5$, and c) $w$ varies quite rapidly at $z<1$ reaching a value of $w>-1$ today (see Figure 2). Finding that $w\ne-1$ would in itself be a huge discovery, but b) and c) are the real curve balls. The most ``simple" physical model for dark energy would be a single scalar field minimally coupled to gravity (e.g., quintessence), which is constrained to lie within the parameter space $-1\leq w \le 1$. Thus, the DESI$+$ results suggest a more complex dark sector than many would have previously assumed \cite[][]{wolf2025, Ye_2025}. However, the evidence for b) is more tentative than for a) and models that relax curvature can also describe the data \cite[][]{dinda2025, chen2025,akrami2025}. The detection of a non-flat universe would also be an astonishing $\Lambda$CDM breaking discovery, but is not discussed further here.

Now reconsider the $H_0$ tension. Although dynamical dark energy describes BAO+CMB+SNe, it does not solve the $H_0$ tension. In other terms, dynamical dark energy fails to fit  BAO+CMB+SNe+local $H_0$. In fact, dynamical dark energy makes the discrepancy in $H_0$ larger! Whereas in $\Lambda$CDM DESI+\textit{Planck} yields $H_0 = 68.17\pm 0.28$ km s$^{-1}$ Mpc$^{-1}$, putting the $H_0$ tension at 5$\sigma$, the best-fit dynamical dark energy model predicts values between $H_0 = 63.6^{+1.6}_{-2.1}$ and $H_0=67.51\pm0.59$ km s$^{-1}$ Mpc$^{-1}$ depending on which assumptions are used. Among the broadest range of precise local measures of $H_0$, none are seen to dip this low \cite{verde2024}. 

Dynamical dark energy in itself is not predicted so much as customized to do the job and may not be the true underlying solution to the observed mis-match between these data sets. It is possible we may find a ``simple" model that fits BAO+CMB+SNe+local $H_0$, and indeed, many theorists are hard at work on this very question!  But it is also possible, especially as error bars shrink, that ``simple" models fail to describe Stage IV data.

The space of potential theoretical models is large (e.g., see Figures 3 and 6  in \cite{cosmoverse}) - here we highlight just a few. Instead of a strongly varying $w$ at late times (such as considered in the DESI papers), ``early dark energy" (EDE) introduces changes to the expansion history around $z=10^4$ to $z=10^3$. The EDE model describes BAO+CMB, and reduces the $H_0$ tension, but is not a perfect fit to SNe data, and also increases the (albeit less significant) ``S$_8$ tension" (see below) \cite{chaussidon2025earlytimesolutionalternative}. Models that consider interactions between dark energy and dark matter have yielded promising results \cite{khoury2025,cruickshank2025,linder2025}. And some are excited by the possibility that the recent DESI results might be explained by string theories which attempt to unify general relativity with quantum mechanics \cite{obied2018sitterspaceswampland, hur2025dynamicaldarkenergydual}. However, the general emerging picture is that simple models (e.g. quintessence) fail to describe the full suite of data including DESI's hints that $w$ may dip below $-1$ and that physical models have yet put all of the required pieces together \cite{linder_interpreting_2024}.

\section{Additional inconsistencies}\label{sectionx}

There are additional inconsistencies in current cosmological data sets. For example, a mismatch in the growth of structure as measured by low redshift probes compared to the CMB,  known as the ``S$_8$ tension". Although recent results from the KiDS survey are now consistent with \textit{Planck} \cite{wright2025kidslegacycosmologicalconstraintscosmic,stolzner2025kidslegacyconsistencycosmicshear}, some S$_8$ tension is observed in other data sets leaving some remaining puzzles \cite{cosmoverse}. Additionally, the gravitational lensing signal around massive galaxies (galaxy-galaxy lensing) and galaxy clustering are mis-matched on small radial ($r\lessapprox 20$ Mpc) scales \cite{leauthaud2017,johannes2023} and the impact of baryons is unexpectedly large \cite{boryana2024}. When analyzed in $\Lambda$CDM, the upper bound on the sum of neutrino masses from DESI is uncomfortably close to constraints from ground based experiments but this tension is eased in a model with dynamical dark energy  \cite{elbers2025}. Another anomaly is the low quadrupole of the CMB power spectrum and related oddities at large angular scales of the temperature fluctuation spectrum \cite[]{2016planck}. Other potential puzzle pieces include the cumulative age of the oldest astrophysical objects and the surprisingly mature appearance of the most distant galaxies as recently seen with JWST \cite{2025jwst}. A longer list of such inconsistencies can be found in \cite{cosmoverse}. These curiosities could be statistical fluctuations, or astrophysical effects, but a subset might also be manifestations of a larger breakdown of $\Lambda$CDM, potentially connected with the same problems as highlighted by $H_0$ and the recent DESI BAO results. 

\section{Looking forward}\label{section6}

Where will new light on this question come from next? With many other Stage IV surveys coming online, we can eagerly anticipate new results within the next few years. First, the recent BAO results are not the final word of DESI on this topic. DESI will have new constraints in 2025-2026 from other probes (e.g., full-shape fitting, bispectrum, gravitational lensing, and peculiar velocities to name a few) using DR2 data. In addition, DESI continues to collect data and new results from BAO with the third data release (DR3) can be expected in 2027. It will be especially interesting to see how the evidence for the phantom crossing evolves with DESI DR3. HST and JWST will continue to sprinkle updates on $H_0$. On the CMB side, the Simons Observatory recently began observations with an initial set of telescopes and could have cosmology results from new temperature and polarization anisotropy measurements as soon as 2026 \cite{Ade_2019}.  The LSST camera recently saw first light at the Vera C. Rubin Observatory with the expectation that the main survey will begin at the end of 2025 \cite{lsstsciencecollaboration2009lsstsciencebookversion}. ESA's  space telescope Euclid launched in 2023 and recently released early data \cite{laureijs2011eucliddefinitionstudyreport}. First  cosmology results from Rubin and Euclid (e.g., gravitational lensing, galaxy cluster counts, supernovae) may come as soon as 2026-2028. The Prime Focus Spectrograph (PFS) on the Subaru Telescope just began its main survey and will focus on BAO measurements at $z>0.8$ \cite{pfs}. 

Beyond these horizons, the $z>1$ universe will be the next frontier for dark energy constraints. NASA has just completed construction of the Nancy Grace Roman 
Space Telescope \cite{spergel2015widefieldinfrarredsurveytelescopeastrophysics}.  Scheduled for launch in 2026, this will be the largest space telescope to take on the hunt for dark sector clues. And finally, the proposed next phase of DESI (``DESI-II") will also push on the $z>1$ frontier in the 2030's \cite{dawson2022snowmass2021cosmicfrontierwhite}. 

If Stage IV surveys break $\Lambda$CDM, what should the approach be moving forward? Are there specific new observational techniques or cross-survey analyses worth developing above others? Or signatures from specific physical models that we should be looking for first? These are not easy questions to answer. Decades from now, we might look back and realize that breaking $\Lambda$CDM was the easy part but that solving the physics of the dark sector was a more complicated story. A post $\Lambda$CDM dark energy task force could help to answer such questions and create a roadmap to guide the path forward.

\section{Discussion}\label{section7}

Has Astronomy’s gamble with dark energy paid off? Have Stage-IV surveys finally started to find evidence of $w\ne-1$? While it is still too early to tell for certain, it looks like the $\Lambda$CDM model is beginning to show cracks at the seams. How would our community digest the possibility of a 5$\sigma$ definitive nail in the $\Lambda$-coffin in the near future?  The history of paradigm shifts in cosmology is marked by conflict, and $\Lambda$CDM may not go gently into the night. But if we break $\Lambda$CDM, what may we learn? We might discover a beautiful new shiny theory, as simple and elegant as the emergence of General Relativity, but we might also find a new cocktail, more potent than $\Lambda$CDM. This cocktail could have multiple types of dark matter, multiple dark energy fields or eras of significance, interactions between the two, and a new relativistic particle to mention a few. An additional breakdown of general relativity on cosmic scales could complicate the picture even further. In this second scenario, we would be a long way off from a physical model and we could be chasing dark sector shadows for many decades to come. Cosmology has generally taken its cues from high energy physics where simplicity is a guiding principle, but if the dark sector shares any of the complexity of the luminesce sector, these assumptions could undermine the hunt. We should be guided by the increasingly robust data—secured through considerable investment—in our pursuit of deeper understanding, rather than settling for descriptive phenomenology. 

Two intentional, cultural changes to the way in which we traditionally practice cosmology could help us prepare for the potential rocky road ahead. First, train and reward our early career scientists to work on complex problems within large teams as opposed to emphasizing the cult of the individual. Second, recognize that new breakthroughs may require complex analyses spanning multiple Stage IV (and ultimately Stage V) data-sets. There are many reasons why we are not set up to do this well (e.g., funding structure, authorship rules, publication policies, differences in codes, lack of support for data releases, or lack of data release at all). Teams and funding opportunities are largely segregated and cross-survey analyses are complex to set up and navigate. Knowing this, we could be ramping up efforts to create the political and structural foundations that would facilitate cross-survey work. Without a clear plan, the beyond-$\Lambda$CDM era could be hap-hazardous and difficult to navigate, especially for early career scientists. At worst, mis-information could emerge from the challenge of replicating and refereeing results. A new road-map similar to the one produced by the dark energy task force could guide the way. Ultimately, the biggest barrier to solving the physics of the dark sector may be the human challenge of coordinating and sustaining effort on complex analyses spanning multiple teams, datasets, funding agencies, international borders, and long time frames.  These challenges are significant, but not insurmountable. Meanwhile, the reward, a deepening understanding of the Universe, is beyond measure.

\section{Acknowledgments}

We thank Eric Linder and Kevin Bundy for reviewing and commenting on this perspective.

\section{Author contributions}

A.L drafted sections 1, 5, 6, 7. A.R drafted sections 2, 3, 4. A.L and A.R both worked on sections 8 and 9.

\section{Competing Interests}

The authors declare no competing interests.


%\bibliography{NatureComment}% common bib file

%\bibliography{}

\end{document}